\documentclass[doublecol]{epl2} 
\usepackage{shinzato-command}
\title{Belief Propagation Algorithm for Portfolio Optimization Problems}
\shorttitle{Belief Propagation Algorithm for Portfolio Optimization Problems} 

\author{Takashi Shinzato\inst{1} \and Muneki Yasuda\inst{2}}
\shortauthor{T. Shinzato \etal}

\institute{                    
  \inst{1} Department of Management 
  Science and Engineering, Graduate School of Systems Science and 
  Technology, Akita Prefectural University, Yurihonjo, Akita, 015-0055, 
  Japan \\
  \inst{2} Department of Applied Information Sciences, 
  Graduate School of Information Sciences,  Tohoku University, Sendai, 
  Miyagi, 980-8579, Japan
}

\abstract{
The typical behavior of optimal 
solutions to portfolio optimization problems with absolute deviation
 and expected shortfall models 
using replica analysis was pioneeringly estimated by S. Ciliberti and M. 
 M$\acute{\rm e}$zard [Eur. Phys. B. {\bf 57}, 175 (2007)]; however, they have not yet
developed an approximate derivation method for finding the optimal 
portfolio with respect to a given return set. 
In this study, an approximation algorithm based on belief propagation 
for the portfolio optimization problem is presented using the Bethe free energy 
formalism, and 
the consistency of the numerical experimental results of the proposed algorithm 
with those of replica analysis is
confirmed. Furthermore, the 
conjecture of H. Konno and H. Yamazaki, that the optimal solutions
 with the absolute deviation model and 
with the mean-variance model have the same typical behavior, is verified using 
replica analysis and the belief propagation algorithm.}

\begin{document}

\maketitle

\section{Introduction}
Portfolio optimization is one of the most fundamental 
frameworks of risk diversification management. Its theory was introduced
by Markowitz in 1959 and is one of the most important areas 
being actively investigated in 
financial engineering\cite{Markowitz,Ciliberti,Konno}. 
In their theoretical research, Ciliberti and M$\acute{\rm e}$zard 
assessed the 
typical behavior of optimal solutions to portfolio optimization problems, 
in particular those described by the absolute deviation and expected shortfall 
models, 
using replica analysis, one of the most powerful approaches in 
disordered systems. With this approach, they showed that the phase 
transitions 
of these optimal solutions were nontrivial\cite{Ciliberti}. 
However, they did not develop 
an 
effective algorithm for finding the optimal portfolio with respect to a 
fixed return set. This requires a rapid algorithm for resolving the 
optimal portfolio problem with respect to a large enough in-sample set.

As a first step in such a research direction,  we propose an algorithm 
based on belief 
propagation, which is well-known as one of the most prominent 
algorithms in probabilistic inference, to resolve the microscopic averages 
of the optimal 
solution in a feasible amount of time 
for a fixed return set. We also confirm whether the numerical experimental 
results of our novel algorithm are consistent with the ones of replica 
analysis. Furthermore, the conjecture of Konno and Yamazaki, that 
if the return at each period is independently and identically drawn from 
the normal probability distribution\cite{Konno}, the optimal portfolio of the 
mean-variance model is consistent with that of the absolute deviation 
model, is supported using replica analysis and belief propagation.


\section{Model Setting}
Let us define the model setting for our discussion. 
A portfolio of $N$ assets and the return at period 
$\mu$ are represented by
$\vec{w}=\left\{w_1,w_2,\cdots,w_N\right\}^{\rm T}\in{\bf R}^N$ and 
$\vec{x}_\mu=\left\{x_{1\mu},x_{2\mu},\cdots,x_{N\mu}\right\}^{\rm 
T}\in{\bf R}^N$, respectively, where $w_k$ is the position of asset $k$, 
and we assume for simplicity that the mean of the return 
of asset $k$ in period $\mu$, $x_{k\mu}$, is zero. 
The notation ${\rm T}$ indicates matrix transposition. 
Given a return set for $p$ periods as reference, the problem is to minimize 
the following cost function (i.e., Hamiltonian) for the portfolio:
\bea
H\left(\vec{w}\right)
\eq\sum_{\mu=1}^pR\left(\f{\vec{w}^{\rm T}\vec{x}_\mu}{\sqrt{N}}
\right),\label{eq1}
\eea
where $R(u)$ represents a cost function, such as $\f{u^2}{2}$ in the mean-variance 
model and 
$|u|$ in the absolute deviation model, respectively. Furthermore, since 
the budget is assumed to be finite, the following global constraint is set:
\bea
\sum_{k=1}^Nw_k\eq N.\label{eq2}
\eea
One of our aims is to develop an effective general algorithm for solving 
this problem; 
in particular, our aim is an algorithm that works for all cost functions 
$R(u)$ and all probability distributions of the returns.

As a basis for the proposed algorithm, following examples in 
statistical mechanics, we set the joint probability of 
portfolio $\vec{w}$ used in Eq.~(\ref{eq1}) using 
finite inverse absolute temperature $\b$ as follows: 
\bea
P(\vec{w})&\propto&
P_0(\vec{w})\exp\left[-\b H\left(\vec{w}\right)\right]\nn
&\propto& \prod_{\mu=1}^p\left[P_0(\vec{w})g\left(\f{\vec{w}^{\rm T}\vec{x}_\mu}{\sqrt{N}}
\right)\right]P_0^{1-p}(\vec{w}),
\eea
where $g(u)=e^{-\b R(u)}$ is the likelihood function and prior 
probability $P_0(\vec{w})\propto \exp\left[\tilde{m}\left(\sum_{k=1}^Nw_k-N\right)
\right]$ for sufficiently large $N$. 
Notice that the 
partition function of this posterior probability 
$Z=\sum_{\vec{w}}\prod_{\mu=1}^p\left[P_0(\vec{w})g\left(\f{\vec{w}^{\rm 
T}\vec{x}_\mu}{\sqrt{N}}\right)\right]P_0^{1-p}(\vec{w})$ is implicitly 
ignored in this analysis because intuitively it is possible to evaluate the 
first- and 
second-order moments of portfolio $w_k$ approximately without the 
partition function by the following procedure.
An arbitrary test probability of portfolio is defined as follows:
\bea
Q\left(\vec{w}\right)&\propto&\prod_{\mu=1}^pb_\mu\left(\vec{w}\right)
\prod_{k=1}^Nb_k^{1-p}(w_k),
\eea
where the reducibility condition on beliefs $b_k(w_k)$ and $b_\mu(\vec{w})$,
\bea
b_k(w_k)\eq\sum_{\vec{w}\setminus w_k}b_\mu\left(\vec{w}\right),\label{eq5}
\eea
must hold and $\vec{w}\setminus w_k$ denotes a subset of $\vec{w}$ 
from which $w_k$ is excluded. The Kullback-Liebler divergence 
(KLD)
$KL(Q|P)=\sum_{\vec{w}}Q(\vec{w})\log\f{Q(\vec{w})}{P(\vec{w})}$ 
provides a useful guideline for deriving the belief propagation 
algorithm. However, since it is too complicated to directly assess KLD except 
in specific graphical models, we here approximate the Bethe free energy denoted as follows:
\bea
F_{\rm Bethe}\eq\sum_{\mu=1}^p\sum_{\vec{w}}
b_{\mu}(\vec{w})
\log\left(\f{b_\mu(\vec{w})}{P_0(\vec{w})g\left(\f{\vec{w}^{\rm T}\vec{x}_\mu}{\sqrt{N}}\right)}
\right)\nn
&&+(1-p)\sum_{k=1}^N\sum_{w_k}b_k(w_k)\log\left(\f{b_k(w_k)}{P_{0k}(w_k)}\right),\label{eq6}
\eea
where $P_{0k}(w_k)\propto e^{\tilde{m}w_k}$ is used. The purpose of 
this step is to derive the optimal portfolio using the beliefs 
$b_k(w_k)$ and $b_\mu(\vec{w})$ that minimize the Bethe free energy under 
the reducibility condition of Eq.~(\ref{eq5}). By adding the term 
$\sum_{\mu=1}^p\sum_{k=1}^N\sum_{w_k}\l_{k\mu}(w_k)\left[\sum_{\vec{w}
\setminus w_k}
b_\mu(\vec{w})-b_k(w_k)\right]$ to the right-hand side of Eq.~(\ref{eq6}), it is 
possible to maximize the Bethe free energy 
with respect to the beliefs to obtain
\bea
b_k(w_k)&\propto&P_{0k}(w_k)\exp\left[\f{1}{1-p}\sum_{\mu=1}^p\l_{k\mu}(w_k)\right],\nonumber\\
b_\mu(\vec{w})&\propto&P_0\left(\vec{w}\right)g\left(\f{\vec{w}^{\rm T}\vec{x}_\mu}{\sqrt{N}}\right)\exp
\left[-\sum_{k=1}^N\l_{k\mu}(w_k)\right].\nonumber
\eea
Furthermore, for simplicity, 
we set 
$\tilde{\l}_{k\mu}(w_k)=\f{1}{1-p}\sum_{\mu=1}^p\l_{k\mu}(w_k)+\l_{k\mu}
(w_k)$ as novel  auxiliary functions, and then 
$b_k(w_k)$ and $b_\mu(\vec{w})$ can be rewritten using 
$\f{1}{1-p}\sum_{\mu=1}^p\l_{k\mu}(w_k)=\sum_{\mu=1}^p\tilde{\l}_{k\mu}
(w_k)$ and $\l_{k\mu}(w_k)=-\sum_{\nu(\ne\mu)}\tilde{\l}_{k\nu}(w_k)
$ as $b_k(w_k)\propto 
P_{0k}(w_k)\exp\left[\sum_{\mu=1}^p\tilde{\l}_{k\mu}(w_k)\right]$ and 
$b_\mu(\vec{w})\propto 
P_0(\vec{w})g\left(\f{\vec{w}^{\rm T}\vec{x}_\mu}{\sqrt{N}}\right)
\exp\left[\sum_{k=1}^N\sum_{\nu(\ne\mu)}\tilde{\l}_{k\nu}(w_
k)\right]$. Moreover, applying the cumulant generating functions
\bea
\phi_k(\theta_k)\eq\log\sum_{w_k}b_k(w_k)e^{w_k\theta_k},\\
\phi_\mu\left(\vec{\theta}\right)\eq\log\sum_{\vec{w}}b_\mu(\vec{w})e^{\vec{w}^{\rm T}\vec{\theta}},
\eea
the first and second moments of $w_k$ have the compact forms 
$m_{wk}=\pp{\phi_k(\theta_k)}{\theta_k}=\pp{\phi_\mu(\vec{\theta})}{\theta_k}$ and 
$\chi_{wk}=\pp{^2\phi_k(\theta_k)}{\theta_k^2}=\pp{^2\phi_\mu(\vec{\theta})}{\theta_k^2}$
 at $\vec{\theta}=\left\{\theta_1,\cdots,\theta_N\right\}^{\rm T}\to0$. This allows us to disregard the calculation 
 of the partition function. 
Then, our proposed algorithm for sufficiently large $N$ comprises the following:
\bea
\label{eq11}m_{wk}\eq \chi_{wk}\left(h_{wk}+\tilde{m}\right),\\
h_{wk}\eq\f{1}{\sqrt{N}}\sum_{\mu=1}^px_{k\mu}m_{u\mu}+
\tilde{\chi}_{wk}m_{wk}\label{eq10},\\
\tilde{\chi}_{wk}\eq\f{1}{N}\sum_{\mu=1}^px_{k\mu}^2\chi_{u\mu},\\
\label{eq11-1}\chi_{wk}\eq\f{1}{\tilde{\chi}_{wk}},\\
\label{eq15}m_{u\mu}\eq\pp
{}{h_{u\mu}}\log\area Dzg\left(z\sqrt{\tilde{\chi}_{u\mu}}+h_{u\mu}\right),\\
h_{u\mu}\eq\f{1}{\sqrt{N}}\sum_{k=1}^Nx_{k\mu}m_{wk}-
\tilde{\chi}_{u\mu}m_{u\mu}\label{eq14},\\
\tilde{\chi}_{u\mu}\eq\f{1}{N}\sum_{k=1}^Nx_{k\mu}^2\chi_{wk},\\
\label{eq18}\chi_{u\mu}\eq-\pp
{^2}{h_{u\mu}^2}\log\area 
Dzg\left(z\sqrt{\tilde{\chi}_{u\mu}}+h_{u\mu}\right),
\eea
where $Dz=\f{dz}{\sqrt{2\pi}}e^{-\f{z^2}{2}}$ is used. Note that if 
$\tilde{\l}_{k\mu}(w_k)$ is redefined as 
$\tilde{\l}_{k\mu}(w_k)=-\f{\g_{k\mu}}{2}w_k^2+\tilde{h}_{k\mu}w_k$, then $\tilde{\chi}_{wk}=\sum_{\mu=1}^p\g_{k\mu}$ and 
$h_{wk}=\sum_{\mu=1}^p\tilde{h}_{k\mu}$ \cite{Kabashima,Minka}. In addition, 
$\tilde{\chi}_{wk}m_{wk}$ and $\tilde{\chi}_{u\mu}m_{u\mu}$ describe the Onsager 
reaction terms in the literature of spin glass theory (respectively\cite{Opper1,Opper2}).

Four points should be noticed here. First, the 
calculation of this procedure is reduced from  $O(N^3)$ to 
$O(N^2)$. For instance, in the case of the mean-variance model, although we 
are required to 
calculate the inverse matrix 
of the correlation 
matrix of return set $XX^{\rm T}\in{\cal M}_{N\times N}$, where return matrix $X=\left\{\vec{x}_1,\cdots,\vec{x}_p\right\}\in{\cal M}_{N\times 
p}$, in 
order to assess the optimal solution rigorously, it is well-known that this 
calculation is $O(N^3)$. Moreover, fortunately it is found that in the case 
of the mean-variance model, this 
algorithm derives the exact optimal solution (see appendix A for details). Second, only Eqs.~(\ref{eq15}) and 
(\ref{eq18}) are dependent on the likelihood function $g(u)=e^{-\b R(u)}$, 
and the variables of index $u$ are the only model dependent ones. 
Furthermore, $\tilde{m}$ is determined by 
Eqs.~(\ref{eq2}) and (\ref{eq11}). 
Third, the randomness of return is not assumed to be sampled 
from specific distributions. Because it is plausible that the assumption 
on the Bethe free energy 
approximation works correctly if the 
return at each period is asymptotically not correlated with other returns . 
Lastly, we expect that in the limit as  $\b\to\infty$, the estimate 
of the portfolio of asset $k$, $m_{wk}$, 
asymptotically corresponds to the optimal portfolio with respect to the 
given return set.
\section{Application}
\begin{figure}
\includegraphics[width=8cm,height=6cm]{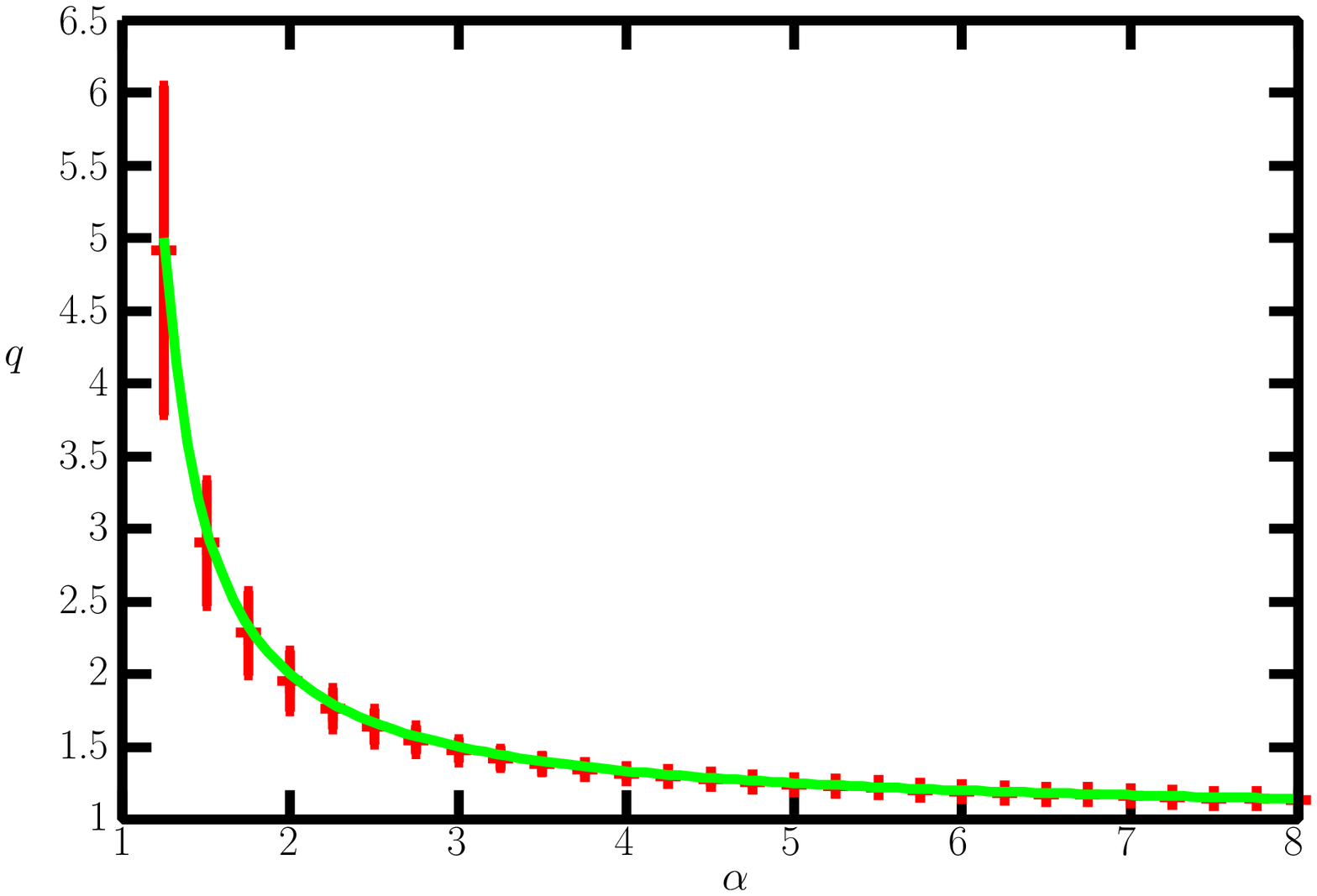}
\caption{The reference ratio $\a=p/N$ (horizontal axis) versus the quenched overlap 
 parameter $q$ (vertical axis). The numerical experimental results from the 
proposed 
 algorithm (error bars) are assessed from $10^2$ 
 experiments using $N=100$ assets. 
Comparing with the results of 
 replica analysis (solid line), the effectiveness of 
 proposed algorithm is verified.
\label{fig.1}}
\includegraphics[width=8cm,height=6cm]{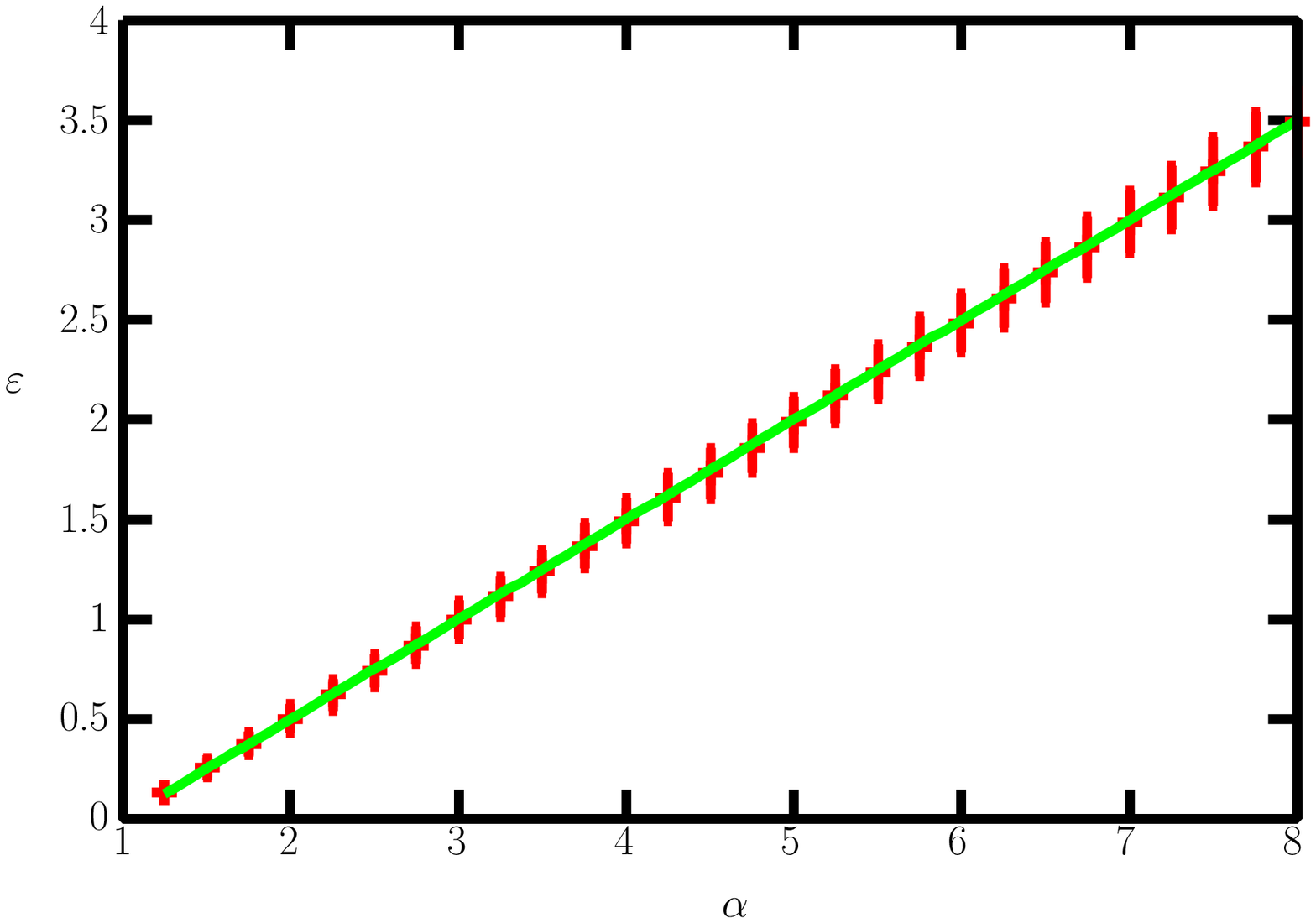}
\caption{The reference ratio $\a$ (horizontal axis) versus one degree of the cost function 
 $\ve$ (vertical axis). This result also indicates that the approximation approach based 
 on probabilistic inference works correctly. \label{fig.2}}
\end{figure}

In order to confirm the effectiveness of our method,
 the numerical experimental results of the proposed algorithm and those 
of the replica analysis for the case of the Markowitz 
model are shown 
in Figs.~\ref{fig.1} and \ref{fig.2}, where $x_{k\mu}$ are independently and 
identically drawn from the normal distribution with mean and 
variance $0$ and $1$, respectively. The numerical experimental result of belief 
propagation is assessed from $10^2$ samples of the number 
of assets $N=100$ and is denoted by error bars and 
 the result of replica analysis is denoted by a solid line. Both findings indicate that the two 
approaches are consistent with each other.

With regard to the conjecture of Konno and Yamazaki, the variables 
in Eqs.~(\ref{eq15}) and 
(\ref{eq18}), in the case of the 
mean-variance model 
\bea
\label{eq17}m_{u\mu}\eq-\f{\b}{1+\b\tilde{\chi}_{u\mu}}h_{u\mu},\\
\label{eq18-1}\chi_{u\mu}\eq\f{\b}{1+\b\tilde{\chi}_{u\mu}}
\eea
and the absolute deviation model
\bea
m_{u\mu}
\eq
\b\tanh\left(\b h_{u\mu}+\f{1}{2}\log\f{H\left(\b\sqrt{\tilde{\chi}_{u\mu}}+\f{h_{u\mu}}{\sqrt{\tilde{\chi}_{u\mu}}}\right)}
{H\left(\b\sqrt{\tilde{\chi}_{u\mu}}-\f{h_{u\mu}}{\sqrt{\tilde{\chi}_{u\mu}}}\right)}
\right),\nn
\\
\chi_{u\mu}
\eq
-\pp{m_{u\mu}}{h_{u\mu}},
\eea
are assessed exactly using $H(u)=\int_u^\infty Dz$. Because $H(u)\simeq \left(\sqrt{2\pi}u\right)^{-1}e^{-\f{u^2}{2}}$ in the case 
of $u\gg1$, $m_{u\mu}\simeq-\f{h_{u\mu}}{\tilde{\chi}_{u\mu}}$ and 
$\chi_{u\mu}\simeq\f{1}{\tilde{\chi}_{u\mu}}$ are estimated; that is, 
this finding indicates that the conjecture of Konno and Yamazaki is 
valid in part in the sense of the belief propagation approach. See appendices for details.

\section{Conclusion}

In conclusion, we have discussed an effective algorithm for finding the optimal solution of the
portfolio optimization problem with respect to an arbitrary cost function 
 according to Ciliberti and M$\acute{\rm e}$zard\cite{Ciliberti}. 
With loss of generality, applying the likelihood function $g(u)$ defined by 
the 
cost function $R(u)$ dependent on the risk diversification problem, we 
proposed a novel approximation derivation method based on 
one of the most powerful estimation methods in probabilistic inference. 
In addition, since two types of Onsager reaction terms are derived in 
Eqs~(\ref{eq10}) and (\ref{eq14}), our 
algorithm provides the Thouless, Anderson, and Palmer approach 
rather than the 
mean-field approximation in the literature of spin glass theory. 
One advantage of  our algorithm is that it rapidly converges by excluding 
the 
effect of self-response.
 In order to confirm the effectiveness of the proposed approach, 
we have described the case of the mean-variance model. 
Furthermore, we have shown that the 
conjecture of Konno and Yamazaki is supported by employing 
both approaches developed in cross-disciplinary research involving 
statistical mechanics and information sciences. In future work, we 
will assess the properties of $R(u)$ and the randomness of return that make solving the portfolio 
optimization problem using belief 
propagation possible.

\acknowledgments
We thank K. Inakawa, Y. Kimura, and K. Yagi for their fruitful comments. 
This work is partially supported by the Grants-in-Aid (No. 21700247) for 
Scientific Research from the Ministry of Education, Culture, Sports, 
Science and Technology of Japan.

\section{Appendix A: Proof of Exactness}
We here confirm the exactness of the proposed belief propagation algorithm 
 for the 
case of the Markowitz model. Our discussion is 
restricted to $\a>1$ for simplicity. 
From Eqs.~(\ref{eq14}), (\ref{eq17}), and 
(\ref{eq18-1}), we obtain $\vec{m}_u=-\f{\b}{\sqrt{N}}X^{\rm T}\vec{m}_w$,
 where   $\vec{m}_u=\left\{m_{u1},\cdots,m_{up}\right\}^{\rm 
T}\in{\bf R}^p$ and $\vec{m}_w=\left\{m_{w1},\cdots,m_{wN}\right\}^{\rm 
T}\in{\bf R}^N$. Furthermore, 
$\tilde{m}\vec{e}=-\f{1}{\sqrt{N}}X\vec{m}_u$ follows  immediately
from Eqs.~(\ref{eq11}), 
(\ref{eq10}), and (\ref{eq11-1}), where 
$\vec{e}=\left\{1,\cdots,1\right\}^{\rm T}\in{\bf R}^N$. 
Thus, substituting $\vec{m}_w=N{\tilde{m}}
\left(\b XX^{\rm T}\right)^{-1}\vec{e}$
into the constraint $N=\vec{e}^{\rm T}\vec{m}_w$ gives the exact optimal solution $\vec{m}_w=\f{N\left(XX^{\rm 
T}\right)^{-1}\vec{e}}{\vec{e}^{\rm T}\left(XX^{\rm 
T}\right)^{-1}\vec{e}}$.

\section{Appendix B: Replica Analysis} 
According to Ciliberti and M$\acute{\rm e}$zard and Varga-Hoszonits and 
Kondor 
\cite{Ciliberti,Varga}, replica symmetry solution of the portfolio optimization 
problem, where $x_{k\mu}$ is independently and identically distributed 
with $N(0,1)$, is derived as the following extremum:
\bea
-\b f\eq\lim_{N\to\infty}\f{1}{N}\left[\log Z\right]_q\nn
\eq\mathop{\rm Extr}_{q,\chi}\left\{\f{q-1}{2\chi}+\f{1}{2}\log\chi
\right.\nn
&&\left.
+\a\area Dy\log
\area Dzg\left(z\sqrt{\chi}+y\sqrt{q}\right)
\right\},\qquad
\eea
where 
$Z=\sum_{\vec{w}}P_0(\vec{w})\prod_{\mu=1}^pg\left(\f{\vec{w}^{\rm 
T}\vec{x}_\mu}{\sqrt{N}}\right)$ is the partition function and the 
notation $\left[\cdots\right]_q$ denotes the quenched average over the return 
set. Moreover, the quenched overlap parameters become 
$q_{ab}=\f{1}{N}\sum_{k=1}^Nw_{ka}w_{kb}=\chi+q$ if $a=b$ and $q$ 
otherwise  by employing replica indices $a,b=1,2,\cdots,n$ and the 
assumption of replica 
symmetry. Furthermore, 
for large $N$ and $p$, $\a=p/N\sim O(1)$ remains finite and plays an 
important role as a control parameter with respect to phase transition phenomena. 
 If 
$g(u)=e^{-\f{\b}{2}u^2}$, then $q=\left(1-\f{1}{\a}\right)^{-1}$ and 
$\chi=\left(\b(\a-1)\right)^{-1}$ can be exactly calculated in the case 
$\a>1$ and $q\to\infty$, and $\chi\to\infty$ otherwise. This analytical 
finding is also verified in by the following. 
It is well known that the eigenvalue distribution of the correlation 
matrix $C=\f{1}{N}XX^{\rm T}$ in the limit of $N\to\infty$ is asymptotically 
close to the Mar$\check{\rm c}$henko-Pastur law
$\rho(\l)=\left[1-\a\right]^+\d(\l)+\f{\sqrt{\left[\l-\l_-\right]^+\left
[\l_+-\l\right]^+}}{2\pi\l}$ with $\l_\pm=\left(1\pm\sqrt{\a}\right)^2$ 
and $\left[u\right]^+=\max\left\{u,0\right\}$ \cite{Shinzato1}. Therefore,
 $q=\left\langle\f{1}{\l^2}\right\rangle 
\left\langle\f{1}{\l}\right\rangle^{-2}$ and one degree of the cost function 
$\ve=\lim_{N\to\infty}\f{1}{N}\left[H(\vec{w})\right]_q=\f{1}{2}\left\langle
\f{1}{\l}\right\rangle^{-1}$ are obtained straightforwardly using $\left\langle 
f(\l)\right\rangle=\area d\l\rho(\l)f(\l)$. Applying 
Mar$\check{\rm c}$hencko-Pastur law, that $\left\langle
\f{1}{\l}\right\rangle=\f{\l_++\l_-}{4\sqrt{\l_+\l_-}}-
\f{1}{2}=\f{1}{\a-1}$ and $\left\langle
\f{1}{\l^2}\right\rangle=\f{\sqrt{\l_+\l_-}}{2\pi}\times
\f{\pi}{2}\left(\f{1}{2}\left(\f{1}{\l_-}-\f{1}{\l_+}\right)\right)^2=
\f{\a}{(\a-1)^3}$ if $\a>1$ and approach infinity 
otherwise follows directly\cite{Kondor,Pafka1,Pafka2,Papp}. This is 
consistent with the findings of replica analysis.

In general, the order parameters are derived as follows:
\bea
\chi\eq-\f{\sqrt{q}}{\a\eta},\label{eq20}\\
q\eq1+\a\chi^2\d,\label{eq21}
\\
\eta\eq\area Dyy\left(\f{\dis{\area Dzg'\left(z\sqrt{\chi}+y\sqrt{q}\right)}}
{\dis{\area Dzg\left(z\sqrt{\chi}+y\sqrt{q}\right)}}\right),\\
\d\eq\area Dy
\left(\f{\dis{\area Dzg'\left(z\sqrt{\chi}+y\sqrt{q}\right)}}
{\dis{\area Dzg\left(z\sqrt{\chi}+y\sqrt{q}\right)}}\right)^2.
\eea
From Eqs. (\ref{eq20}) and (\ref{eq21}),
\bea
q\eq\left(1-\f{1}{\a}\f{\d}{\eta^2}\right)^{-1}
\eea
is obtained. In the limit of sufficiently large $\b$ of 
$g(u)=e^{-\b|u|}$, if we assess $\eta\simeq-\f{\sqrt{q}}{\chi}$ and 
$\d\simeq\f{q}{\chi^2}$ asymptotically, then the conjecture of Konno 
and Yamazaki is confirmed as correct in the sense of 
replica analysis.

\section{Appendix C: The Conjecture of Konno and Yamazaki}
This conjecture is related to the assessment of an annealed system in 
the context of spin glass theory. 
If the return at period $\mu$, $\vec{x}_\mu$, is independently and identically 
drawn from $N\left(0,\Sigma\right)$, where $\Sigma\in{\cal M}_{N\times 
N}$ is variance-covariance matrix and $\vec{w}$ is fixed, the novel 
variable $z=\f{\vec{w}^{\rm 
T}\vec{x}_\mu}{\sqrt{N}}$ is distributed as 
$N\left(0,s^2(\vec{w})\right)$ with 
$s^2(\vec{w})=\f{1}{N}\vec{w}^{\rm T}\Sigma\vec{w}\in{\bf R}$. 
With respect to fixed $\vec{w}$, employing one 
degree of the cost function of the 
annealed optimization problem 
$\ve(\vec{w})=\left[\f{1}{N}\sum_{\mu=1}^pR\left(\f{\vec{w}^{\rm 
T}\vec{x}_\mu}{\sqrt{N}}\right)\right]_q=\a\area 
DuR\left(us(\vec{w})\right)$, which becomes 
$\ve_{\rm MV}(\vec{w})=\f{\a}{2}s^2(\vec{w})$ in 
the case of the mean-variance model and $\ve_{\rm 
AD}(\vec{w})=\f{2\a}{\sqrt{2\pi}}|s(\vec{w})|$ 
in the absolute deviation model. 
This implies that the optimal portfolios of the 
annealed situations of the two models are consistent with each other. Note 
that one degree of the cost function in the case 
of the annealed portfolio problem with the expected shortfall model, $\ve_{\rm 
ES}(\vec{w})=\min_{v\ge0}\a\left\{v\g+H\left(\f{v}{s(\vec{w})}\right)
\right\}$ with $\g>0$ can also be assessed\cite{Rockafellar}. 
If $s(\vec{w})\le\f{1}{\sqrt{2\pi}\g}$, 
then this optimal solution is identical to 
those of the previous mentioned models. 
This finding, that is, 
$\arg\mathop{\min}_{\vec{w}^{\rm T}\vec{e}=N}\ve_{\rm MV}(\vec{w})
=\arg\mathop{\min}_{\vec{w}^{\rm T}\vec{e}=N}\ve_{\rm AD}(\vec{w})
$, is one part of the contributions reported by Konno and 
Yamazaki.

However, they optimistically assumed 
$\vec{w}_{\rm MV}=\vec{w}_{\rm AD}$ 
with respect to a given return set $X$ without any mathematical 
proof, using 
\bea
\vec{w}_{\rm MV}\eq\arg\mathop{\min}_{\vec{w}^{\rm T}\vec{e}=N}
\f{1}{2N}\sum_{\mu=1}^p\sum_{i=1}^N\sum_{k=1}^Nw_iw_kx_{i\mu}x_{k\mu},\\
\vec{w}_{\rm AD}\eq\arg\mathop{\min}_{\vec{w}^{\rm T}\vec{e}=N}
\sum_{\mu=1}^p\left|\f{1}{\sqrt{N}}\sum_{k=1}^Nw_kx_{k\mu}\right|.
\eea
As explained above, $\arg\mathop{\min}_{\vec{w}^{\rm T}\vec{e}=N}\ve_{\rm MV}(\vec{w})
=\arg\mathop{\min}_{\vec{w}^{\rm T}\vec{e}=N}\ve_{\rm AD}(\vec{w})
$ with respect to the annealed optimization problem strictly holds; 
however, $\vec{w}_{\rm MV}=\vec{w}_{\rm AD}$ is not always satisfied. For example, 
in the simple case of $N=p=2$ for the two returns $\vec{x}_1=\left\{a,c\right\}^{\rm T}$ and 
$\vec{x}_2=\left\{b,d\right\}^{\rm T}$, their assumption 
$\vec{w}_{\rm MV}=\vec{w}_{\rm AD}$ does not 
hold, except under specific special situations.

Although this is apparently contradictory to these obtained findings 
from both approaches, it is necessary to recognize that the relation 
$\vec{w}_{\rm MV}=\vec{w}_{\rm AD}$ with a fixed return set is 
equivalent to  the sufficient 
condition $q_{\rm MV}=q_{\rm AD}$, where 
$q_{\rm MV}=\lim_{N\to\infty}\f{1}{N}
\left[{\vec{w}_{\rm MV}^{\rm 
T}\vec{w}_{\rm MV}}
\right]_q$ and $q_{\rm AD}=\lim_{N\to\infty}\f{1}{N}
\left[{\vec{w}_{\rm AD}^{\rm T}\vec{w}_{\rm AD}}\right]_q
$ are quenched averages of overlap parameters.
Moreover, although $\vec{w}_{\rm MV}=\vec{w}_{\rm AD}$ does not hold 
in general, 
it is expected that 
the inner product 
$\f{\vec{w}_{\rm MV}^{\rm T}\vec{w}_{\rm AD}}{\left|\vec{w}_{\rm 
MV}\right|\left|\vec{w}_{\rm AD}\right|}$ is approximately 1
because $\f{1}{N}\sum_{\mu>\nu}x_{k\mu}x_{j\nu}\to0$ in the case of 
sufficiently large $N$\cite{Gulyas}.

\end{document}